\documentclass{emulateapj}

\usepackage{natbib}
\usepackage{amsmath}

\newcommand{\about}{$\sim\!\!$~}
\newcommand{\kms}{\,km\,s$^{-1}$}

\def\lsim{\hbox{\rlap{\raise 0.425ex\hbox{$<$}}\lower 0.65ex\hbox{$\sim$}}}
\def\gsim{\hbox{\rlap{\raise 0.425ex\hbox{$>$}}\lower 0.65ex\hbox{$\sim$}}}

\def\arcdeg{\mbox{$^\circ$}}

\newcommand{\mean}[1]{\left \langle #1 \right \rangle}

\shorttitle{SN~Ia Ejecta Velocity and Intrinsic Color}
\shortauthors{Foley \& Kasen}

\begin{document}

 \title{Measuring Ejecta Velocity Improves Type~Ia Supernova
 Distances}

\def\cfa{1}
\def\clay{2}
\def\berk{3}
\def\lbl{4}

\author{
{Ryan~J.~Foley}\altaffilmark{\cfa,\clay} and
{Daniel~Kasen}\altaffilmark{\berk,\lbl}
}

\altaffiltext{\cfa}{
Harvard-Smithsonian Center for Astrophysics,
60 Garden Street, 
Cambridge, MA 02138.
}
\altaffiltext{\clay}{
Clay Fellow. Electronic address rfoley@cfa.harvard.edu .
}
\altaffiltext{\berk}{
Department of Physics,
University of California at Berkeley,
366 LeConte,
Berkeley, CA 94720.
}
\altaffiltext{\lbl}{
Nuclear Science Division,
Lawrence Berkeley National Laboratory,
Berkeley, CA 94720.
}

\begin{abstract}
We use a sample of 121 spectroscopically normal Type Ia supernovae
(SNe~Ia) to show that their intrinsic color is correlated with their
ejecta velocity, as measured from the blueshift of the \ion{Si}{2}
$\lambda 6355$ feature near maximum brightness, $v_{\rm Si~II}$.  The
SN~Ia sample was originally used by \citet{Wang09:2pop} to show that
the relationship between color excess and peak magnitude, which in the
absence of intrinsic color differences describes a reddening law, was
different for two subsamples split by $v_{\rm Si~II}$ (defined as
``Normal'' and ``High-Velocity'').  We verify this result, but find
that the two subsamples have the same reddening law when extremely
reddened events ($E(B-V) > 0.35$~mag) are excluded.  We also show that
(1) the High-Velocity subsample is offset by \about 0.06~mag to the
red from the Normal subsample in the $(B_{\rm max} - V_{\rm
max})$--$M_{V}$ plane, (2) the $B_{\rm max} - V_{\rm max}$ cumulative
distribution functions of the two subsamples have nearly identical
shapes, but the High-Velocity subsample is offset by \about 0.07~mag
to the red in $B_{\rm max} - V_{\rm max}$, and (3) the bluest
High-Velocity SNe~Ia are \about 0.10~mag redder than the bluest Normal
SNe~Ia.  Together, this evidence indicates a difference in intrinsic
color for the subsamples.  Accounting for this intrinsic color
difference reduces the scatter in Hubble residuals from 0.190~mag to
0.130~mag for SNe~Ia with $A_{V} \lesssim 0.7$~mag.  The scatter can
be further reduced to 0.109~mag by exclusively using SNe~Ia from the
Normal subsample.  Additionally, this result can at least partially
explain the anomalously low values of $R_{V}$ found in large SN~Ia
samples.  We explain the correlation between ejecta velocity and color
as increased line blanketing in the High-Velocity SNe~Ia, causing them
to become redder.  We discuss some implications of this result, and
stress the importance of spectroscopy for future SN~Ia cosmology
surveys, with particular focus on the design of WFIRST.
\end{abstract}

\keywords{supernovae: general --- distance scale --- dust, extinction}

\defcitealias{Wang09:2pop}{W09}
\defcitealias{Kasen07:asym}{KP07}


\section{Introduction}\label{s:intro}

Type Ia supernovae (SNe~Ia) are very good distance indicators after
making an empirical correction based on their light-curve shape and
color \citep{Phillips93, Riess96}.  The distances are precise enough
to determine that the expansion of the Universe is accelerating
\citep{Riess98:Lambda, Perlmutter99}, and constrain the
equation-of-state parameter of Dark Energy \citep{Astier06,
Wood-Vasey07, Riess07, Hicken09:de, Kessler09, Amanullah10}.  The
current best precision is 0.15 and 0.11~mag in absolute magnitude
(after corrections) when using optical information only and combining
it with near-infrared data, respectively (Mandel et al., in prep.),
although most samples have a precision of \about 0.17~mag
\citep[e.g.,][]{Hicken09:lc}.

Since peak luminosity, light-curve shape, intrinsic color, and most
spectral parameters all correlate with each other and with measured
$^{56}$Ni mass \citep[e.g.,][]{Nugent95, Stritzinger06}, measuring the
light-curve shape of a SN~Ia is adequate to determine its peak
luminosity.  Observing a SN~Ia in multiple passbands allows one to
estimate its reddening.  Therefore with only light curves in a few
passbands, one can precisely measure the distance to a SN~Ia.

However, as mentioned above, the precision in measuring the distance
modulus of an individual SN~Ia has a floor of 0.15~mag when observing
(rest-frame) optical passbands.  There have been many attempts to
further reduce the Hubble scatter with a ``second parameter'' that
does not strongly correlate with other luminosity indicators.  In
particular, since spectra must contain more information than light
curves, several investigations have looked for correlations between
spectral parameters and both light-curve shape and Hubble residuals,
with most correlations between spectral parameters and light-curve
shape not improving the Hubble scatter.  However, two studies have
been able to make a further reduction to the Hubble scatter using the
ratios of fluxes at particular wavelengths \citep{Foley08:uv,
Bailey09}, but both require further data to verify the results and a
theoretical understanding of why particular wavelengths have such
power.

\citet{Benetti05} showed that although a single parameter can
adequately account for the variation in the peak luminosity,
light-curve shape, and temperature of SNe~Ia, the velocity gradient of
the \ion{Si}{2} $\lambda 6355$ feature, $\dot{v}_{\rm Si~II}$, is not
well correlated with these other observables.  The \ion{Si}{2}
$\lambda 6355$ feature, being relatively strong and isolated in
maximum-light spectra, is the hallmark feature of SNe~Ia \citep[for a
review of SN spectroscopy, see][]{Filippenko97}.  \citet{Benetti05}
separated their sample of SNe~Ia into three groups: high-velocity
gradient (HVG) objects, consisting of SNe~Ia with $\dot{v}_{\rm Si~II}
\gtrsim 70$~km~s$^{-1}$~day~$^{-1}$ and $\Delta m_{15} (B) \lesssim
1.5$~mag, low-velocity gradient (LVG) objects, consisting of SNe~Ia
with $\dot{v}_{\rm Si~II} \lesssim 70$~km~s$^{-1}$~day~$^{-1}$ and
$\Delta m_{15} (B) \lesssim 1.5$~mag, and FAINT objects which have
$\Delta m_{15} (B) \gtrsim 1.5$~mag.  The HVG and LVG subsamples have
similar light-curve shape, peak luminosity, and color demographics.
Finally, \citet{Benetti05} found that HVG SNe~Ia tended to have a
higher velocity for the \ion{Si}{2} feature near maximum than the
LVG SNe~Ia.

\citet{Wang09:2pop}, hereafter \citetalias{Wang09:2pop}, presented an
analysis of 158 spectroscopically normal SNe~Ia (i.e.,
``Branch-normal'' objects; \citealt{Branch93:normal}; which excludes
SNe~Ia similar to SN~1991T; \citealt{Filippenko92:91T};
\citealt{Jeffery92}; SN~1991bg; \citealt{Filippenko92:91bg};
\citealt{Leibundgut93}; and more peculiar SNe~Ia; e.g.,
SN~2000cx; \citealt{Li01:00cx}; SN~2002cx; \citealt{Li03:02cx};
SN~2003fg; \citealt{Howell06}; SN~2008ha; \citealt{Foley09:08ha}),
which were separated by their \ion{Si}{2} $\lambda 6355$ velocity near
maximum light, $v_{\rm Si~II}$.  They showed that this selection was
essentially equivalent to using $\dot{v}_{\rm Si~II}$ for the SNe~Ia
where there was a good spectroscopic sequence.  SNe~Ia above and below
a velocity of \about 11,800~\kms\ were classified as ``High-Velocity''
and ``Normal'' SNe~Ia\footnote{We use ``Normal'' to describe these
objects since \citetalias{Wang09:2pop} used this terminology.  But we
caution that it could be confused with ``Branch normal,'' which both
Normal and High-Velocity SNe~Ia are considered.  Calling these objects
``Low-Velocity'' in the future may reduce this potential confusion.},
respectively.  They looked at the correlation between color excess
(essentially an offset from maximum-light $B-V$ color) and peak
absolute magnitude corrected for light-curve shape, but not for color.
If the color excess is truly the result of dust, this relationship is
between color excess and extinction, and the slope of a line fit to
the relationship should provide a measurement of the ratio of total to
selective extinction, $A_{V} / E(B-V) = R_{V}$.  The objects within
the two subsamples follow a different relation between peak absolute
magnitude (again, uncorrected for color) and color excess, indicating
different values of $R_{V}$ for the two subsamples: $R_{V} = 2.36$ and
1.37 for the Normal and High-Velocity subsamples, respectively.  Both
of these values, are significantly less than the Milky Way value of
$R_{V} = 3.1$.

\citet{Kasen07:asym}, hereafter \citetalias{Kasen07:asym}, presented
theoretical light curves and spectra of a SN~Ia model from several
different viewing angles.  The ejecta in the model were asymmetric
such that different viewing angles had different values for
$\dot{v}_{\rm Si~II}$ and $v_{\rm Si~II}$ but similar values of
$\Delta m_{15} (B)$ and peak magnitude.  Since it was the same
explosion model, all generated light curves and spectra come from an
event with a single total energy and $^{56}$Ni mass.  Although the
details of the model presented by \citetalias{Kasen07:asym} may not
properly describe SNe~Ia, the generic attributes of the observables
are useful when discussing differences in ejecta velocity for SNe~Ia
with similar peak luminosity and/or light-curve shape.

Although neither study focused on this result, both
\citetalias{Wang09:2pop} and \citetalias{Kasen07:asym} found
correlations between $v_{\rm Si~II}$ and the $B-V$ color of the SN at
maximum light.  \citetalias{Wang09:2pop} found a \about 0.1~mag offset
between $B-V$ colors of the Normal and High-Velocity subsamples.
\citetalias{Kasen07:asym} showed that both $B-V$ color and $v_{\rm
Si~II}$ depend on the viewing angle of their asymmetric
two-dimensional model.

This manuscript examines the correlation of color with ejecta
velocity.  It attempts to explain the physical cause of the
correlation and examines the implications for measuring SN~Ia
distances.  We suggest that a single maximum-light spectrum will
significantly improve the precision of the measured distance for a
given SN~Ia.  In Section~\ref{s:wang}, we re-analyze the
\citetalias{Wang09:2pop} sample, examining in detail the relationship
between these quantities.  We find that the intrinsic color of a SN~Ia
at maximum brightness depends on the velocity of the ejecta.  We also
show that a proper treatment of the intrinsic colors of SNe~Ia with
different ejecta velocities can significantly improve the precision of
distance measurements.  In Section~\ref{s:models}, we provide a simple
theoretical explanation for our results as well as examine more
complex models.  In Section~\ref{s:conc}, we summarize our results and
discuss their implications for future studies.


\section{Ejecta Velocity and Intrinsic Color}\label{s:wang}

As discussed above, \citetalias{Wang09:2pop} presented an analysis of
158 SNe~Ia that concluded that separating the objects into two
subsamples based on $v_{\rm Si~II}$ and treating their extinction
corrections differently will reduce the scatter of their Hubble
residuals.  In this section, we re-analyze the data presented by
\citetalias{Wang09:2pop}, showing that their analysis stands.
However, (1) a different $R_{V}$ for the two subsamples is only
necessary when including the reddest events, (2) for SNe~Ia with
$E(B-V) \le 0.3$~mag, both subsamples have reddening laws consistent
with $R_{V} = 3.1$, (3) there is a clear offset in the color of SNe~Ia
in the two subsamples, (4) separating the objects into two subsamples
can partially account for the low $R_{V}$ values measured for large
samples of low-reddening SNe~Ia, and (5) using this difference in
color, one can significantly reduce the scatter of the Hubble
residuals for these SNe~Ia.

From the entire \citetalias{Wang09:2pop} sample, we remove SN~2006bt,
which has spectral and photometric peculiarities that result in a peak
luminosity fainter than its light-curve shape would suggest
\citep{Foley10:06bt}.  Like \citetalias{Wang09:2pop}, we remove
SN~2006lf, which has a Milky Way reddening of $E(B-V) = 0.97$~mag and
was \about 0.7~mag brighter than that of a typical SN~Ia after
light-curve shape correction.  To avoid large systematic uncertainties
for the peak absolute magnitudes, \citetalias{Wang09:2pop} excluded
all SNe with $z < 0.01$ that do not have a Cepheid distances from
their final analysis.  We also remove these objects from our sample.
After making these cuts, 121 SNe~Ia remain in the final sample, with
81 and 40 SNe~Ia being in the Normal and High-Velocity subsamples,
respectively.

\subsection{Verification of Previous Analysis}\label{ss:ver}

To verify the results of \citetalias{Wang09:2pop}, we examine their
sample of SNe~Ia in the same manner as \citetalias{Wang09:2pop}.  In
particular, we use Equation~1 of \citetalias{Wang09:2pop}:
\begin{equation}\label{e:wang}
  M_{\rm max}^{V} = M_{\rm zp} + \alpha (\Delta m_{15} (B) - 1.1) + R_{V} E(B-V)_{\rm host}.
\end{equation}
This two-component fit is similar to the two-component fit used by
\citet{Tripp98} and in the SALT \citep{Guy05, Guy07} and SiFTO
\citep{Conley08} light-curve fitters.

\begin{figure}
\begin{center}
\epsscale{1.}
\rotatebox{90}{
\plotone{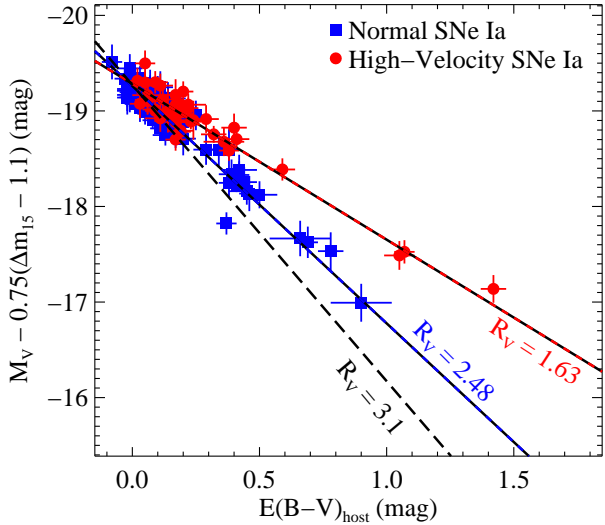}}
\caption{Our re-creation of Figure~4 of \citetalias{Wang09:2pop}.  The
light-curve shape corrected peak absolute $V$ brightness as a function
of host color excess.  The blue squares and red circles represent the
Normal and High-Velocity SNe~Ia, respectively.  The best fit lines for
each subsample correspond to $R_{V} = 2.48$ and 1.63 for the Normal
and High-Velocity SNe~Ia (corresponding to blue-black and red-black
dashed lines), respectively.  The Milky Way reddening law of $R_{V} =
3.1$ is also shown as the dashed line.}\label{f:wang1}.
\end{center}
\end{figure}

\citetalias{Wang09:2pop} provides all necessary information in their
Table~1.  Fitting the data, we find the same values for $M_{\rm zp}$,
$\alpha$, and $R_{V}$ as \citetalias{Wang09:2pop}: $-19.26$~mag, 0.77,
and 2.36; and $-19.28$~mag, 0.75, and 1.58 for the Normal and
High-Velocity SNe~Ia, respectively.  After fixing $\alpha = 0.75$
(similar to \citetalias{Wang09:2pop}), we refit the data, finding
slightly different values for $R_{V}$ ($2.48 \pm 0.11$ and $1.63 \pm
0.08$ for Normal and High-Velocity SNe~Ia, respectively).  These
values differ very slightly (and within the uncertainties) from those
found by \citetalias{Wang09:2pop}.  Figure~\ref{f:wang1} presents our
recreation of Figure~4 from \citetalias{Wang09:2pop}.

From Figure~\ref{f:wang1}, it is clear that the subsamples have
different relationships between color excess and peak absolute
magnitude after correcting for light-curve shape.  Similarly, both
subsamples have reddening laws different from that of the Milk Way.
However, a closer examination of the data is required.

\subsection{Excluding the Reddest Supernovae}\label{ss:wang2}

As noted by \citetalias{Wang09:2pop}, the reddening laws for the two
groups are significantly different even if SNe~Ia with $E(B-V)_{\rm
host} > 0.5$~mag are excluded from the fit.  However, SNe~Ia with
$E(B-V)_{\rm host} = 0.5$~mag are highly reddened objects
(corresponding to $0.8 \le A_{V}\le 1.6$~mag depending on $R_{V}$).
These SNe~Ia are very rarely seen at high redshift and only
occasionally observed in the local universe.  In fact, the only
high-$z$ SN~Ia in the Constitution set of SNe~Ia \citep{Hicken09:de}
with $E(B-V) > 0.5$~mag is from the Higher-$z$ {\it HST} survey
\citep{Riess07} and has $E(B-V)_{\rm host} = 0.52$~mag.  This object
is also at $z = 0.216$, making it the lowest-redshift SN~Ia of the
{\it HST}-discovered SNe~Ia.  For current cosmological analyses, this
color limit is not particularly useful.

\begin{figure}
\begin{center}
\epsscale{1.}
\rotatebox{90}{
\plotone{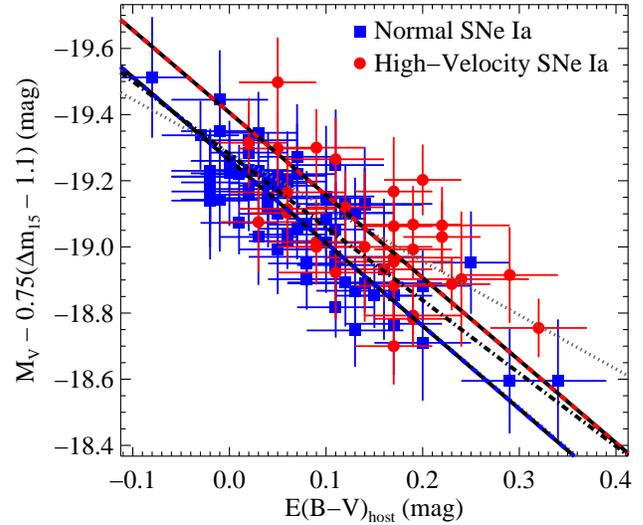}}
\caption{Same as Figure~\ref{f:wang1}, except for SNe~Ia with
$E(B-V)_{\rm host} < 0.35$~mag.  The best-fit lines for the restricted
subsamples are shown as blue-black and red-black dashed lines and
correspond to $R_{V} = 2.51$ and 2.50 for the Normal and High-Velocity
SNe~Ia, respectively.  The best-fit lines for the full subsamples (as
shown in Figure~\ref{f:wang1}) are shown as grey dotted lines and
correspond to $R_{V} = 2.48$ and 1.63 for the Normal and High-Velocity
SNe~Ia, respectively; however, the line for the Normal subsample is
almost completely covered by the line fit to the low reddening Normal
SNe~Ia.  A fit to the full sample of SNe~Ia with $E(B-V)_{\rm host} <
0.35$~mag is shown as a dot-dashed line and corresponds to $R_{V} =
2.21$.}\label{f:wang2}.
\end{center}
\end{figure}

In Figure~\ref{f:wang2}, we again present the data of
\citetalias{Wang09:2pop}, except now we only show the data with
$E(B-V)_{\rm host} < 0.35$~mag (corresponding to $A_{V} < 0.57$, 0.87,
and 1.09~mag for $R_{V} = 1.63$, 2.48, and 3.1, respectively).  For
all further analysis, the light-curve shape parameter, $\alpha$, is
fixed at 0.75.  As a check, we have also performed all analysis
leaving this as a free parameter, and it is always consistent with
0.75.  However, for subsamples with a small number of SNe~Ia the
uncertainty for the parameter can be large, and fixing the parameter
for all fits makes comparing samples more direct and convenient.  For
the subsample with $E(B-V)_{\rm host} < 0.35$~mag, the data produces
best-fit values of $(M_{\rm zp}, R_{V}) = (-19.26 \pm 0.03 {\rm ~mag},
2.51 \pm 0.28)$ and $(-19.41 \pm 0.08 {\rm ~mag}, 2.50 \pm 0.47)$ for
Normal and High-Velocity SNe~Ia, respectively.  The fit parameters for
the low-reddening Normal subsample are consistent with those founds
for the entire Normal subsample.  Conversely, the low-reddening
High-Velocity subsample has values of $M_{\rm zp}$ and $R_{V}$ that
differ at the 1.5 and $1.8 \sigma$ level from those found for the
entire High-Velocity subsample.  Although these differences are not
statistically significant, the 0.12~mag offset in absolute magnitude
and 0.87 offset in $R_{V}$ (corresponding to a difference of 0.17~mag
in $A_{V}$ for $E(B-V) = 0.2$~mag) can dramatically affect the implied
distances of SNe~Ia at cosmological distances.

\begin{figure}
\begin{center}
\epsscale{1.6}
\rotatebox{90}{
\plotone{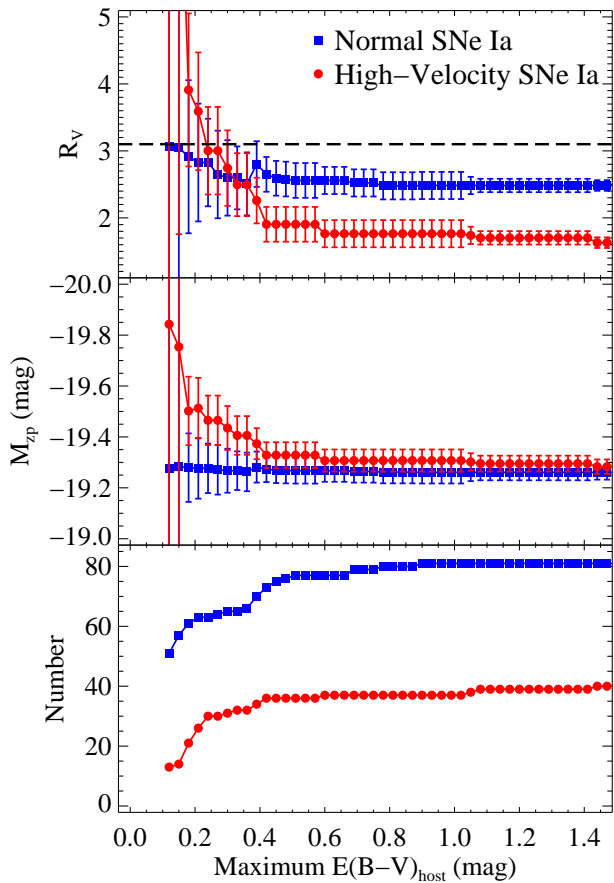}}
\caption{Best-fit $R_{V}$ (first panel) and $M_{\rm zp}$ (second panel),
as well as the number of SNe in each subsample (third panel) as a
function of maximum $E(B-V)_{\rm host}$.  The Normal and High-Velocity
SNe~Ia are represented by the blue squares and red circles,
respectively.  The dashed line in the first panel represents $R_{V} =
3.1$.}\label{f:rv}.
\end{center}
\end{figure}

We further examine this difference by fitting all data with
$E(B-V)_{\rm host}$ below a certain ceiling and varying that maximum
value.  The resulting best-fit parameters are shown in
Figure~\ref{f:rv}.  Naturally, the uncertainties for $M_{\rm zp}$ and
$R_{V}$ are much larger for a smaller $E(B-V)_{\rm max}$.  This is
both because of a smaller number of SNe~Ia in each subsample and the
smaller lever arm for fitting the line.  Regardless, the two
subsamples always have consistent values of $M_{\rm zp}$; however,
they are only consistent at the \about $1 \sigma$ level for
$E(B-V)_{\rm max} < 0.4$~mag.  (Section~\ref{ss:color} shows that this
is likely an offset in color rather than in peak absolute magnitude.)
The two subsamples also have consistent values for $R_{V}$ for
$E(B-V)_{\rm max} < 0.4$~mag, but deviate significantly for higher
values \citepalias{Wang09:2pop}.  Additionally, both subsamples are
consistent with $R_{V} = 3.1$ (at $1 \sigma$) for $E(B-V)_{\rm max}
\le 0.3$~mag.

The fit parameters do not change significantly for any subsample of
Normal SNe~Ia.  But $R_{V}$ changes significantly depending on the
maximum reddening for the High-Velocity SNe~Ia.  It is possible that
the highly reddened High-Velocity SNe~Ia have a value of $R_{V}$ that
is significantly smaller than that of less reddened High-Velocity
SNe~Ia or any Normal SN~Ia.  A low value for $R_{V}$ has been measured
for individual High-Velocity SNe~Ia \citep[e.g.,][]{Krisciunas00,
Wang08:06x}.  We note that SN~2003cg, which is defined as a Normal
object in the \citetalias{Wang09:2pop} sample, but is excluded from
our analysis because of its redshift, was measured to have $R_{V} =
1.80$ \citep{Elias-Rosa06}.  \citet{Goobar08} suggested that a low
value for $R_{V}$ may be the result of multiple scatterings off
circumstellar dust.  In this scenario, it is natural that low values
of $R_{V}$ are associated with large values of $E(B-V)$.  Although the
possibility of circumstellar dust for all High-Velocity SNe~Ia cannot
strictly be ruled out, the higher value of $R_{V}$ for the
low-reddening sample suggests that it is not the dominant component of
any extinction.  Furthermore, as seen in Figure~\ref{f:wang2}, there
is an offset between the Normal and High-Velocity subsamples in the
$M_{V}$--$E(B-V)_{\rm host}$ plane that must be the result of some
linear combination of $M_{V}$ and $E(B-V)_{\rm host}$.  Therefore, the
subsamples have intrinsically different light-curve shape-corrected
peak luminosities (with the High-Velocity subsample being more
luminous), have intrinsically different colors (with the High-Velocity
subsample being redder), and/or intrinsically different dust
distributions (with the High-Velocity subsample having more dust along
the line of sight).  The highly reddened SNe~Ia probably have
different properties from the low-reddening SNe~Ia, with the highly
reddened High-Velocity SNe~Ia having a lower $R_{V}$ than the
low-reddening members of that subsample.  Further investigations are
necessary to determine (1) why highly reddened High-Velocity SNe~Ia
have a low value for $R_{V}$, and (2) if low-reddening SNe~Ia have a
value of $R_{V}$ that is inconsistent with 3.1.

The results of restricting the samples by reddening show that the
extremely reddened objects highly influence the fit parameters.
Depending on choices made to cull the sample or (particularly for
high-$z$ SNe) selection effects, these values may vary dramatically.

\subsection{Color Offsets}\label{ss:color}

By restricting the \citetalias{Wang09:2pop} sample to have
$E(B-V)_{\rm host} < 0.35$~mag, we found in Section~\ref{ss:wang2}
that the two velocity-selected subsamples have different magnitude
zeropoints, $M_{\rm zp}$.  However, this is only one way to interpret
the data; the other possibility is that there is an offset in
$E(B-V)_{\rm host}$ for the two subsamples.  This is effectively
shifting the lines in Figure~\ref{f:wang2} left-right rather than
up-down.  From the best-fit lines, the offset is \about 0.06~mag.

To clearly see this effect, we look again at the data from
\citetalias{Wang09:2pop}.  Instead of using the derived $E(B-V)_{\rm
host}$ values from \citetalias{Wang09:2pop}, we use their values of
$B_{\rm max}-V_{\rm max}$ (in most cases, there is a simple offset
between the two values).  This pseudo-color is simply the $B$
magnitude at $B_{\rm max}$ minus the $V$ magnitude at $V_{\rm max}$,
with no correction for host reddening.  We also restrict the sample to
SNe~Ia with $1 \le \Delta m_{15} (B) \le 1.5$~mag and $B_{\rm max} -
V_{\rm max} < 0.5$~mag.  As noted by \citetalias{Wang09:2pop}, the two
subsamples have similar demographics over this range of $\Delta m_{15}
(B)$.

\begin{figure}
\begin{center}
\epsscale{1.3}
\rotatebox{90}{
\plotone{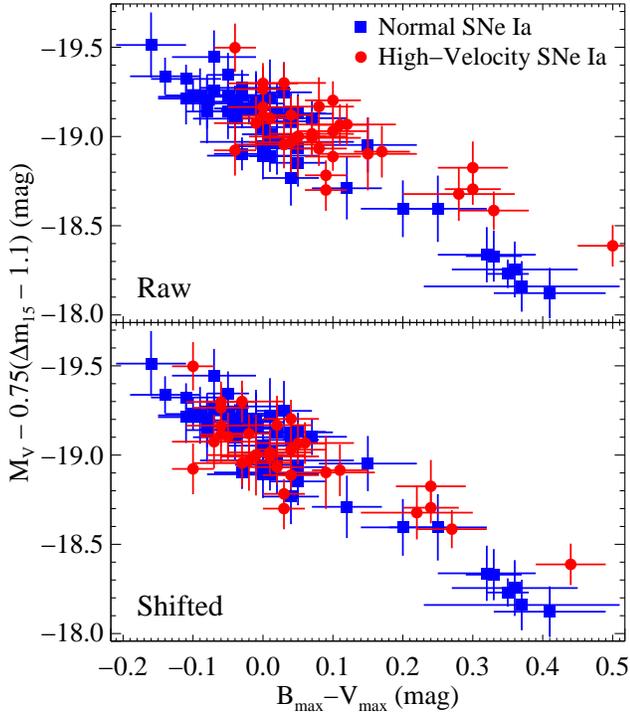}}
\caption{Same as Figure~\ref{f:wang2}, except data are plotted as a
function of $B_{\rm max}-V_{\rm max}$ and only SNe~Ia with $1 \le
\Delta m_{15} (B) \le 1.5$~mag are displayed.  The top panel shows the
raw data, while the bottom panel displays the same data with the
High-Velocity subsample shifted by $B_{\rm max}-V_{\rm max} =
-0.06$~mag.}\label{f:offset}.
\end{center}
\end{figure}

\citetalias{Wang09:2pop} noted that for $B_{\rm max}-V_{\rm max} <
0.2$~mag there was a 0.08~mag offset (0.10~mag if restricting the
$\Delta m_{15} (B)$ range) in the average $B_{\rm max}-V_{\rm max}$
between the two subsamples.  There is also a lack of very blue
($B_{\rm max} - V_{\rm max} < -0.05$~mag) SNe~Ia in the High-Velocity
subsample.  The bluest objects in a sample represent the lowest
reddening and intrinsically bluest SNe~Ia for that sample.  For the
Normal subsample to have several SNe~Ia bluer than the bluest
High-Velocity SN~Ia, {\it all} High-Velocity SNe~Ia must be
significantly reddened and/or the High-Velocity and Normal SNe~Ia have
intrinsically different colors.  In the top panel of
Figure~\ref{f:offset}, Figure~\ref{f:wang1} is remade, except each
SN~Ia is plotted as a function of $B_{\rm max}-V_{\rm max}$ instead of
$E(B-V)_{\rm host}$ and only show SNe~Ia with $1 \le \Delta m_{15} (B)
\le 1.5$~mag.  In the bottom panel of the figure, the same data are
presented, except the High-Velocity SNe~Ia are shifted by $B_{\rm
max}-V_{\rm max} = -0.06$~mag, the value found from the best-fit
lines.

Although by eye the overlap between the Normal and High-Velocity
objects may not appear to be significantly different between the raw
and shifted data, it is clear that the shifted data are a much better
match to the bluest and reddest Normal SNe~Ia than the raw data.
That is, this simple shift adequately accounts for the lack of blue
objects in the High-Velocity subsample and removes the overall offset
in the two populations.

\begin{figure}
\begin{center}
\epsscale{1.}
\rotatebox{90}{
\plotone{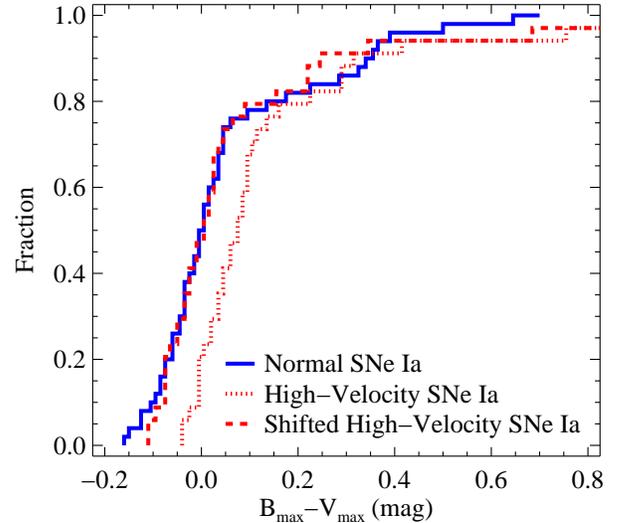}}
\caption{$B_{\rm max}-V_{\rm max}$ CDF for the Normal (blue solid
line), raw High-Velocity (red dotted line), and High-Velocity shifted
by $-0.07$~mag (red dashed line) subsamples.  All SNe~Ia shown here
have $1 \le \Delta m_{15} (B) \le 1.5$~mag.}\label{f:cdf}.
\end{center}
\end{figure}

Another way to examine the effect of shifting the data is by examining
the cumulative distribution functions (CDFs) for the subsamples.  The
CDFs for both subsamples are presented in Figure~\ref{f:cdf}.
Clearly, there is a dearth of blue SNe~Ia in the High-Velocity
subsample, with 26\% of the Normal subsample being bluer than the
bluest High-Velocity SN~Ia.  However, the overall shape of the CDFs
for the Normal and High-Velocity subsamples are similar.  After
shifting the High-Velocity subsample by $-0.07$~mag, the CDFs are very
similar except perhaps for the reddest part of the sample where dust
extinction must be the dominant source of color differences.  A simple
Kolmogorov-Smirnov (K-S) test shows that there is a 0.65\% chance that
the Normal and High-Velocity objects have the same parent population.
The Normal and shifted High-Velocity subsamples have a K-S probability
of 99.99\%, which does not mean that the subsamples have the same
probability (especially since the High-Velocity colors were
arbitrarily shifted by a fixed amount), but does show the outstanding
agreement in the two subsamples' CDFs after the shift.  The value of
$-0.07$~mag was chosen to maximize the K-S probability, but values of
$-0.06$ and $-0.08$~mag yield values of 73.07 and 98.96\%,
respectively.

As discussed in \citetalias{Wang09:2pop}, the offset in maximum-light
colors is likely the result of either different amounts of dust or
intrinsically different colors.  If the offset is primarily from dust,
then the subsamples must have intrinsically different light-curve
shape-corrected peak luminosities.  Additionally, for the CDFs to have
the same shape, but be offset in color, the High-Velocity SNe~Ia must
have an additional amount of dust corresponding to $E(B-V)_{\rm host}
\approx 0.07$~mag for all values of $B_{\rm max}-V_{\rm max}$.
Alternatively, a difference in intrinsic color accounts for the offset
in the CDFs and does not require a difference in the light-curve
shape-corrected peak luminosities for the two subsamples.  The lack of
High-Velocity SNe~Ia with colors as blue as the bluest Normal SNe~Ia
is additional evidence that there is almost certainly an intrinsic
color difference.

\subsection{Properly Measuring $R_{V}$ for a Sample of Supernovae}

Several groups have noted that the Hubble residuals of SNe~Ia are
minimized by choosing a very low (and potentially unphysical) value of
$R_{V}$ \citep[e.g.,][]{Tripp98, Astier06, Guy07, Conley07, Kessler09,
Hicken09:de,Amanullah10}.  The methods vary in detail, but all are
essentially equivalent to fitting a line to the data in a plot like
Figure~\ref{f:wang1} or \ref{f:wang2}.  If SNe~Ia have intrinsically
different colors, this method ignores critical information.

An example of this problem is shown in Figure~\ref{f:wang2}.  We fit
the entire sample of SNe~Ia with $E(B-V)_{\rm host} \le 0.35$~mag from
\citetalias{Wang09:2pop} with a single line.  Since the Normal
subsample has bluer objects for the same $M_{V}$ than the
High-Velocity subsample, this process will naturally reduce the slope
of the fit --- effectively reducing the value of $R_{V}$.  Where both
the Normal and High-Velocity subsamples are best-fit with $R_{V} =
2.50$ separately, combined they are best fit by $R_{V} = 2.21 \pm
0.21$.

If one properly separates these groups, the situation becomes the same
as described in Section~\ref{ss:wang2}.  This clearly affects the
distances measured in previous cosmological analyses and should be
accounted for in the future.

\subsection{Improvement to Measured Distances}\label{ss:distances}

\citetalias{Wang09:2pop} noted that using a different value of $R_{V}$
for the two subsamples would reduce the scatter of Hubble residuals
for their sample.  However, as Section~\ref{ss:wang2} showed, the two
subsamples have similar values of $R_{V}$ if the extremely reddened
SNe~Ia are excluded, and it is not clear that two values of $R_{V}$ is
particularly useful for cosmological samples.  In this section, we
examine how the Hubble scatter is affected by separating the SNe~Ia
into two groups based on their ejecta velocity.

\begin{figure}
\begin{center}
\epsscale{1.}
\rotatebox{90}{
\plotone{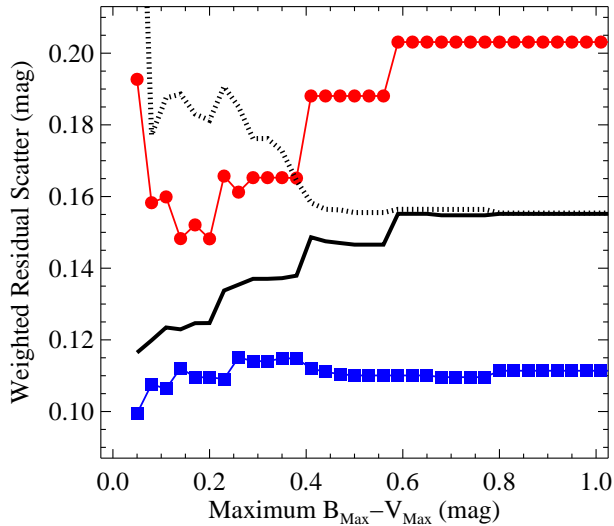}}
\caption{Weighted Hubble residual scatter as a function of maximum
$B_{\rm max} - V_{\rm max}$ color.  The dotted and solid lines
represent the full sample (with $1 \le \Delta m_{15} (B) \le 1.5$~mag)
assuming a single intrinsic color and separate intrinsic colors for
Normal and High-Velocity SNe~Ia, respectively.  The blue squares and
red circles represent the Normal and High-Velocity subsamples,
respectively.  Using separate intrinsic colors for the two subsamples
significantly reduces the scatter.  Additionally, Normal SNe~Ia
appear to have significantly smaller scatter than High-Velocity
SNe~Ia.}\label{f:hubble}.
\end{center}
\end{figure}

Figure~\ref{f:hubble} presents the weighted residual scatter, defined
as
\begin{equation}
  \left ( \frac{\sum \left ( \Delta \mu^{2} / \sigma^{2} \right )}
    {\sum \left ( 1 / \sigma^{2} \right )} \right )^{1/2},
\end{equation}
where $\Delta \mu$ and $\sigma$ are the Hubble residual the
uncertainty of the distance modulus for a given SN~Ia, respectively,
for different samples as a function of maximum $B_{\rm max} - V_{\rm
max}$ color.  In the figure, the full sample (with $1 \le \Delta
m_{15} (B) \le 1.5$~mag) is represented by the black lines, while the
Normal and High-Velocity subsamples are represented by blue squares
and red circles, respectively.  The fits which generate the black
dotted line assume a single intrinsic color for the entire sample, and
allow $M_{\rm zp}$ and $R_{V}$ to vary for each maximum value of
$B_{\rm max} - V_{\rm max}$.  This is the approach that has normally
been used in cosmological analyses.  To determine the scatter for the
other samples, all parameters were fixed for all maximum values of
$B_{\rm max} - V_{\rm max}$, and only a color offset between the
Normal and High-Velocity SNe~Ia was applied.

\begin{figure}
\begin{center}
\epsscale{1.}
\rotatebox{90}{
\plotone{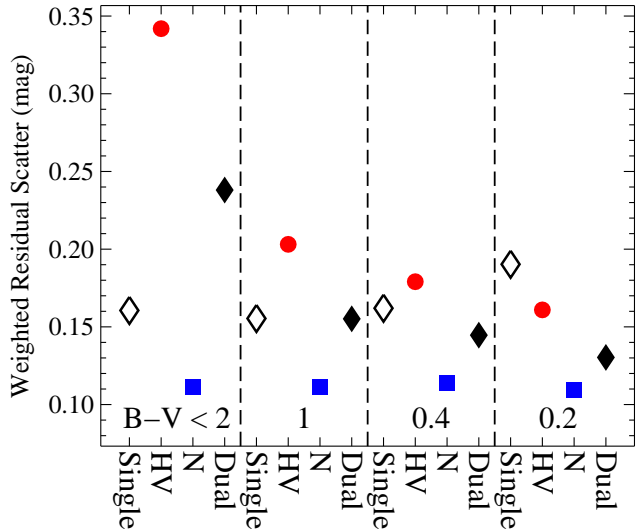}}
\caption{Weighted Hubble residual scatter as a function of maximum
$B_{\rm max} - V_{\rm max}$ color for various samples and methods.
The empty and filled black diamonds represent the full sample (with $1
\le \Delta m_{15} (B) \le 1.5$~mag) assuming a single intrinsic color
(labeled `Single') and separate intrinsic colors (labeled `Dual') for
Normal and High-Velocity objects, respectively.  The blue squares and
red circles represent the Normal (labeled `N') and High-Velocity
(labeled `HV') subsamples, respectively.  Only the values for $B_{\rm
max} - V_{\rm max} \le 0.2$, 0.4, 1, and 2~mag (from right to left)
for each sample and method are plotted.}\label{f:hubble2}.
\end{center}
\end{figure}

For $B_{\rm max} - V_{\rm max} \le 1$~mag, the scatter is consistently
improved by using two intrinsic colors.  For SNe~Ia with $B_{\rm max}
- V_{\rm max} \le 0.2$~mag (corresponding to $E(B-V) \lesssim 0.3$~mag
and $A_{V} \lesssim 0.7$~mag), the scatter decreases from 0.190~mag to
0.130~mag by adopting this method.  Furthermore, the Normal subsample
has a scatter of only 0.109~mag for this color cut.  This is better
visualized in Figure~\ref{f:hubble2}, where only a subset of the data
from Figure~\ref{f:hubble} is plotted.

Increasing the maximum color increases the scatter of the
High-Velocity subsample, but the scatter of the Normal subsample
remains constant for all color ranges.  This is a consequence of the
Normal subsample having a single value for $R_{V}$ for all colors
while the High-Velocity SNe~Ia have a significantly different value
for $R_{V}$ depending on the maximum color (see Section~\ref{ss:wang2}
and Figure~\ref{f:rv}).  This, in turn, increases the scatter of the
full sample for large values of $B_{\rm max} - V_{\rm max}$, even when
using different intrinsic colors for the subsamples.  Perhaps
counterintuitively, the scatter of the full sample with a single
intrinsic color increases as the maximum color is decreases.  This is
mostly because the fraction of Normal to High-Velocity SNe~Ia
increases with increasing color in our sample.

Despite the relatively small number of High-Velocity SNe in this
sample, it is noteworthy that the Normal SNe have a significantly
better Hubble residual scatter --- {\it even after treating each
subsample separately}.  A future study may find a color parameter that
scales with velocity, rather than a simple splitting of the sample,
will help improve this measurement.  But in the era of large SN
samples, where systematic uncertainties are larger than statistical
uncertainties and the sample sizes are large enough to choose the
``best'' objects, using only Normal SNe~Ia for measuring cosmological
parameters may be a prudent choice.  Of course it goes without saying
that such a selection (at least at this time) would require a
spectrum.

Our analysis is relatively simplistic.  Peak absolute magnitudes are
corrected for light-curve shape using only a linear relationship with
$\Delta m_{15} (B)$, the reddening estimate is derived essentially
only from $B_{\rm max} - V_{\rm max}$, and the distance moduli are
derived from peak absolute magnitudes in a single band ($V$).
Furthermore, the SN~Ia sample is split by velocity rather than having
a model with velocity as a parameter (this is currently necessary
since \citetalias{Wang09:2pop} has not released their $v_{\rm Si~II}$
measurements).  Nonetheless, this initial analysis provides the
foundation for more sophisticated analyses.  In particular, $v_{\rm Si
II}$ (or perhaps a more general ejecta velocity parameter) appears to
be a true ``second parameter'' which significantly improves the
precision with which we can measure distances to SNe~Ia beyond that of
just light-curve shape and color.


\section{Theoretical Understanding}\label{s:models}

Thus far, empirical relationships between ejecta velocity and
intrinsic color have been presented.  In this section, we present a
simple explanation for these correlations.  We then suggest that this
scenario is a natural prediction of asymmetric SN explosions.
Finally, we show that a set of model light curves and spectra also
show a correlation between color and ejecta velocity.

\subsection{A Toy Model}

An extremely simplistic model of a SN SED near maximum brightness is
the following: a blackbody spectrum is generated from electron
scattering in the ejecta, imprinted on this spectrum are broad P-Cygni
lines from strong lines, and this causes significant absorption at
wavelengths shorter than \about 4300~\AA, where line-blanketing from
Fe-group elements is significant.  The wavelength where line opacity
and electron-scattering opacity are similar is at \about 4300~\AA\
\citep{Kasen07:wlr}, which is near the peak of the $B$ band.

As ejecta velocity increases, one expects the width of line-forming
regions in velocity space to increase (i.e., at higher velocities,
lines get broader); this is especially true for saturated lines.
Therefore, individual lines overlap more for higher-velocity ejecta.
For wavelengths dominated by electron scattering, this should not
significantly affect the SED.  However, for regions where line opacity
is dominant, or in regions where the two effects are of similar
strength, higher velocities should increase the opacity.  Since the
$B$ band is a region where the line and electron-scattering opacities
are of similar strength, the $B$-band flux should decrease with higher
velocity.  However, the $V$ band, which is in a region dominated by
electron scattering, will be less affected by higher ejecta velocity,
and the $V$-band flux should not change significantly.  Therefore, one
{\it expects} that higher ejecta velocity should result in a redder
$B-V$ color.

SN ejecta have a negative temperature gradient with radius.  Since the
outer layers are cooler, SNe with higher velocities will have more
absorption from lower excitation (e.g., \ion{Fe}{2} vs.\ \ion{Fe}{3})
lines.  Because of the larger number of optical lines for the lower
excitation states, higher-velocity ejecta should produce additional
line blanketing at blue wavelengths.  Again, one {\it expects} that
higher ejecta velocity should result in a redder $B-V$ color.

This has been shown in some theoretical
models. \citetalias{Kasen07:asym} generated light curves and spectra
for a single asymmetric SN~Ia explosion model from different lines of
sight.  The asymmetry of the \citetalias{Kasen07:asym} model is the
result of a surface detonation, but the final ejecta distribution is
qualitatively similar to that of general off-center explosions,
resulting either from off-center ignition or detonation conditions, or
both \citep[e.g.,][]{Maeda10:asym}.  The asymmetry of the model is not
(at this point) important for the discussion, but it allows an
examination of SNe~Ia with different ejecta velocities for the same
explosion energy.  There is a clear correlation between viewing angle
and both $v_{\rm Si~II}$ and $B-V$ at maximum light.  Although this
was not explicitly mentioned by
\citetalias{Kasen07:asym}, the different lines of sight showed that
with increasing $v_{\rm Si~II}$ the $B-V$ color became redder.  This
is a validation of the simple explanation presented above.

There are several ways to generate different ejecta velocities for a
given luminosity, but this is also a natural outcome of an asymmetric
explosion, where different viewing angles have similar luminosities
but vastly different ejecta velocities.  Multi-dimensional theoretical
models of SN~Ia explosions have shown that they can be highly
asymmetric \citep[e.g.,][]{Reinecke02, Gamezo05, Kuhlen06}.  Further
theoretical models have shown that the asymmetry can significantly
affect observables depending on viewing angle \citep{Kasen09}.  From
this work, it is expected that SNe~Ia be asymmetric and that the
asymmetry to increase the dispersion of SN distances, which in turn
decreases the precision with which SNe~Ia can constrain cosmological
parameters.

Spectropolarimetry of SNe~Ia has shown that most SNe~Ia have very low
continuum polarization, but can have significant line polarization,
especially in the \ion{Si}{2} $\lambda 6355$ feature (e.g.,
\citealt{Leonard05}; \citealt{Wang07}; see \citealt{Wang08:specpol}
for a review).  Spectropolarimetry probes asymmetries of the SN in the
plane of the sky.  The data suggest that the global asymmetry of a
SN~Ia is typically $\lesssim 10\%$, but that a large plume of Si-rich
material with a preferred axis may be present in all SNe~Ia.
\citet{Wang07} also presented tantalizing evidence that the
light-curve shape of a SN~Ia is correlated with the amount of
polarization in the \ion{Si}{2} feature.

Recently, \citet{Maeda10:neb} found that there were significant
offsets from zero velocity for relatively unblended forbidden lines in
the nebular spectra of SNe~Ia.  These lines probe the inner ejecta of
the SN, and if the explosion were symmetric, one would expect them to
be centered at zero velocity.  \citet{Maeda10:asym} found a striking
correlation between the velocity offsets of these features and the
velocity gradient of the \ion{Si}{2} $\lambda 6355$ feature,
$\dot{v}_{\rm Si~II}$, at early times.  This observation connects the
outer layers of the ejecta to the explosion near the center of the
star, and suggests an intrinsically asymmetric explosion mechanism for
the majority of SNe~Ia.  \citet{Maund10} also found a correlation
between the amount of polarization in the \ion{Si}{2} feature and
$\dot{v}_{\rm Si~II}$, suggesting that $\dot{v}_{\rm Si~II}$ is an
excellent probe of asymmetry of the outer layers of the SN ejecta.

Our work also fits within this framework.  Intrinsic color correlates
with $v_{\rm Si~II}$, which correlates with $\dot{v}_{\rm Si~II}$.
Since $\dot{v}_{\rm Si~II}$ does not correlate with luminosity, we
require two explosions with similar luminosities to have vastly
different ejecta velocities.  That is naturally explained if SNe~Ia
are asymmetric with all lines of sight having a similar luminosity but
some lines of sight showing significantly different ejecta velocities
than others.

\subsection{Model Light Curves \& Spectra}\label{ss:models}

\begin{figure}
\begin{center}
\epsscale{1.25}
\rotatebox{90}{
\plotone{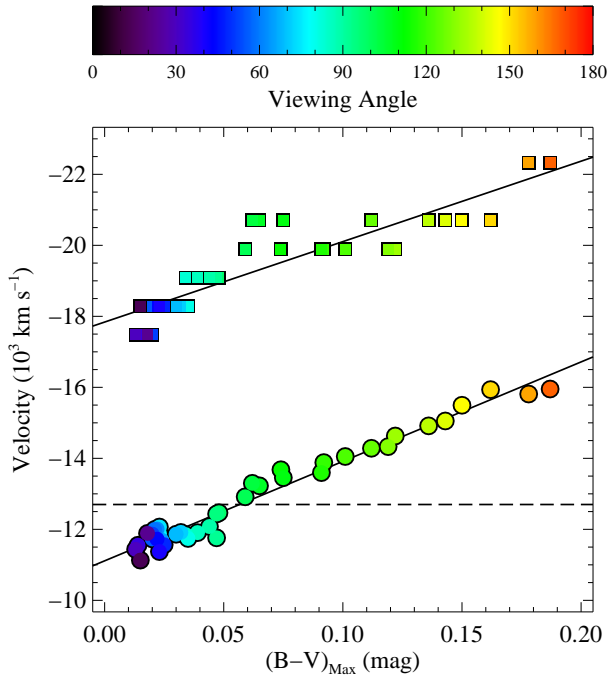}}
\caption{Maximum-light \ion{Si}{2} $\lambda 6355$ (circles) and Ca
H\&K (squares) velocity as a function of maximum-light $B-V$ color for
the \citetalias{Kasen07:asym} model spectra.  Each viewing angle is
represented by a different color with the mapping shown by the color
bar at the top.  The best-fit lines for each feature are shown as
black solid lines.  The dashed line represents the $v_{\rm Si~II}$
offset from the mean of the Normal velocity clump that
\citetalias{Wang09:2pop} used to differentiate the Normal and
High-Velocity subsamples.  It also roughly separates viewing angles
into two hemispheres.}\label{f:bv_vel}.
\end{center}
\end{figure}

The \citetalias{Kasen07:asym} models can be used to examine
correlations between various observables with the hope of
understanding the physics behind our empirical findings.
\citetalias{Kasen07:asym} already noted that $B-V$ color evolution and
$v_{\rm Si~II}$ depend on viewing angle.  However, they did not
examine the maximum-light values of these two quantities and their
correlation.  Figure~\ref{f:bv_vel} shows the $(B-V)_{\rm max}$ color
and $v_{\rm Si~II}$ as a function of viewing angle.  The two values
are highly correlated, and a linear fit produces
\begin{equation}
  v = \left ( -11.1 - 28.0 (B-V)_{\rm max} \right ) \times 10^{3} {\rm ~km~s}^{-1}.
\end{equation}

A similar relationship exists for the Ca H\&K feature, with
\begin{equation}
  v = \left ( -17.8 - 22.7 (B-V)_{\rm max} \right ) \times 10^{3} {\rm ~km~s}^{-1}.
\end{equation}
Not surprisingly, the velocity of the Si and Ca features are highly
correlated in the models.  However, unlike the \ion{Si}{2} $\lambda
6355$ feature, which was fit with a Gaussian to determine the minimum
of the feature, the Ca H\&K feature had a very complex profile for
some viewing angles, and the velocity was simply measured from the
minimum of the profile.  Thus the precision of this measurement is
limited by the resolution of the model spectra.  Real SN~Ia data show
complex, non-Gaussian line profiles with multiple absorption minima
for the Ca H\&K feature \citep[e.g.,][]{Branch06}.  This is further
complicated by varying quality for each observed spectrum.  Future
work should determine if Ca H\&K correlates with color in a manner
similar to \ion{Si}{2}.

While our toy model and the models of \citetalias{Kasen07:asym}
predict that $B_{\rm max} - V_{\rm max}$ monotonically increases with
ejecta velocity, our current analysis simply splits the sample into
two groups based on $v_{\rm Si II}$.  Future studies should determine
from the data if a linear relationship is more appropriate.  This
simplification may also be why the Hubble scatter is larger for the
High-Velocity subsample than the Normal subsample.  Additionally, the
velocity used to separate the samples is somewhat arbitrary and there
could be some blending between the samples, leading to an increase in
the Hubble residual scatter.  A linear relationship would avoid such a
potential bias.

\citetalias{Wang09:2pop} found that the Normal SNe~Ia had a
mean value of $\mean{v_{\rm Si II}} \approx -10,600$~km~s$^{-1}$, and
a value of $v_{\rm Si II} \approx -11,800$~km~s$^{-1}$ was used to
split the sample into the two subsamples.  Using the same velocity
difference from the mean to separate the \citetalias{Kasen07:asym}
model spectra (the lower-velocity clump is at $v_{\rm Si II} \approx
-11,500$~km~s$^{-1}$, so the separation would be at $v_{\rm Si II}
\approx -12,700$~km~s$^{-1}$), the velocity separation also cleanly
separates the models into a lower-velocity clump and a higher-velocity
tail.  Interestingly, this separation also almost perfectly separates
the viewing angles into two hemispheres.  The
\citetalias{Kasen07:asym} model colors are also systematically redder
than the observations, but this trend is typical of other light curve
models \citep[e.g.,][]{Kasen09, Kromer09} and is likely due to the
neglect of non-local thermodynamic equilibrium effects in the
radiation transport calculations.

Though the analysis here was done with a single aspherical model, the
same sorts of effects can been seen in spherical models when the
explosion energy is varied for a given $^{56}$Ni mass.  We have
examined the survey of parameterized 1-D models of \citet{Woosley07}
and confirmed that a higher kinetic energy results in a redder color
at peak.


\section{Discussion \& Conclusions}\label{s:conc}

We have re-analyzed a large sample of SN~Ia data originally presented
by \citetalias{Wang09:2pop}.  We have verified the results of
\citetalias{Wang09:2pop}: after splitting the sample by $v_{\rm
Si~II}$, the two subsamples have different relationships between the
light-curve shape-corrected (but not host-galaxy extinction-corrected)
peak magnitude and color excess, and that difference can be explained
with different reddening laws for the subsamples.  However, we find
that this conclusion no longer holds when the reddest SNe~Ia are
excluded from the sample, and in this regime, both subsamples can be
described by the same reddening law.  Furthermore, the two subsamples
are offset from each other in the plane of color and absolute
magnitude.  Since the bluest High-Velocity SNe~Ia are significantly
redder than the bluest Normal SNe~Ia (also noticed by
\citetalias{Wang09:2pop}) and the CDFs of the two subsamples have the
same shape, but are also offset in color, we conclude that the two
subsamples have different intrinsic colors.

This result has significant implications for SN~Ia cosmology.  First,
by assuming a single intrinsic color for all SNe~Ia, one will
naturally find a smaller value for $R_{V}$ than if the subsamples are
treated separately.  Second, assuming a single intrinsic color could
significantly bias cosmological results, particularly if the fraction
of Normal to High-Velocity SNe~Ia changes with redshift, which may
have already been seen in data \citep{Kessler09}.  Finally, we find
that the scatter in the Hubble residuals for SNe~Ia with $B_{\rm max}
- V_{\rm max} \le 0.2$~mag is reduced from 0.190~mag to 0.130~mag by
using two separate intrinsic colors for the velocity subsamples.  The
Hubble scatter is further reduced to 0.109~mag by exclusively using
the Normal subsample.  To perform this measurement, a single spectrum
within a week of maximum brightness is required.

The correlation between ejecta velocity and color can naturally be
explained by additional line blanketing for the higher-velocity
SNe~Ia, affecting the SED in the $B$ band, but leaving the $V$ band
not as greatly affected.  We show that a simple asymmetric explosion
model can reproduce the general trends in the data.

The relationship between ejecta velocity and intrinsic color will be a
significant tool for measuring extremely precise distances with
SNe~Ia.  Better theoretical understanding of this relationship and
additional observational investigations are necessary to fully
leverage this finding.  It is also important to incorporate this
information into existing SN~Ia distance estimators, which currently
depend solely on light-curve information.  We have only investigated
color near maximum-brightness, and it is important to see how ejecta
velocity correlates with color at all phases --- especially at late
phases when SNe~Ia are assumed to have the same intrinsic color
regardless of light-curve shape \citep{Lira98}.  If future surveys
want the most precise SN~Ia distances possible, they should have a
significant spectroscopic component.  This is particularly important
for the design of WFIRST, where a decision must be made regarding
whether a spectrograph (and with what resolution) should be included.

\begin{acknowledgments} 

\bigskip
R.J.F.\ is supported by a Clay Fellowship.

R.J.F.\ would like to thank the many people with whom he discussed
this study.  In particular, R.\ Kirshner, P.\ Mazzali, M.\ Phillips,
S.\ Jha, K.\ Mandel, S.\ Hachinger, W.\ Li, X.\ Wang, and F.\
R\"{o}pke, and S.\ Blondin influenced the direction of this work.
Part of the analysis occurred at the Aspen Center for Physics during
the Summer 2010 workshop, ``Taking Supernova Cosmology into the Next
Decade.''  Additional analysis occurred while visiting the
Max-Planck-Intitut f\"{u}r Astrophysik.  R.J.F.\ would especially like
to thank W.\ Hillebrandt for his support during this stay.  The
environment that he, along with P.\ Mazzali, F.\ R\"{o}pke, and the
numerous postdoctoral and graduate students have generated is
extremely stimulating.

\end{acknowledgments}

\bibliographystyle{fapj}
\bibliography{astro_refs}


\end{document}